\documentclass[preprints,article,accept,moreauthors,a4paper,pdftex]{Definitions/mdpi}
\pdfoutput=1

\firstpage{1}
\pubvolume{x}
\issuenum{x}
\articlenumber{x}
\pubyear{2021}
\copyrightyear{2021}
\history{Date: 27 May 2021}

\usepackage{accents}
\usepackage{float}
\newcommand{\lc}[1]{\accentset{\circ}{#1}}
\newcommand{\weff}{\mathrm{w}_{\mathrm{eff}}{}}

\usepackage{amsmath}
\usepackage{amsfonts}
\usepackage{amssymb}
\usepackage{ulem}

\usepackage{subfigure}

\Title{Global Portraits of Nonminimal Teleparallel Inflation}

\Author{Laur Järv $^{1}$ and Joosep Lember $^{1}$}

\AuthorNames{Laur Järv, Joosep Lember}

\address{%
$^{1}$ \quad Laboratory of Theoretical Physics, Institute of Physics,
University of Tartu, W. Ostwaldi 1, 50411 Tartu, Estonia}

\corres{Correspondence: laur.jarv@ut.ee}

\abstract{We construct the global phase portraits of inflationary dynamics in teleparallel gravity models with a scalar field nonminimally coupled to torsion scalar. The adopted set of variables can clearly distinguish between different asymptotic states as fixed points, including the kinetic and inflationary regimes. The key role in the description of inflation is played by the heteroclinic orbits which run from the asymptotic saddle points to the late time attractor point and are approximated by nonminimal slow roll conditions. To seek the asymptotic fixed points we outline a heuristic method in terms of the ``effective potential'' and ``effective mass'', which can be applied for any nonminimally coupled theories.
As particular examples we study positive quadratic nonminimal couplings with quadratic and quartic potentials, and note how the portraits differ qualitatively from the known scalar-curvature counterparts. For quadratic models inflation can only occur at small nonminimal coupling to torsion, as for larger coupling the asymptotic de Sitter saddle point disappears from the physical phase space. Teleparallel models with quartic potentials are not viable for inflation at all, since for small nonminimal coupling the asymptotic saddle point exhibits weaker than exponential expansion, and for larger coupling disappears too.}


\begin{document}

\section{Introduction}

While early introductions of a nonminimal coupling between the scalar field and curvature were motivated by e.g.\ Mach's principle \cite{Brans:1961sx} or  conformal invariance \cite{Chernikov:1968zm}, the nonminimal coupling appears naturally as a result of quantum corrections to the scalar field on curved spacetime \cite{Birrell:1982ix} as well as in the effective actions of higher dimensional constructions \cite{Overduin:1998pn,Nastase:2019mhe}. The scalar-tensor framework also offers a convenient representation of many other gravitational theories, like $f(R)$ \cite{Clifton:2011jh,Capozziello:2011et}. On the phenomenological side the nonminimal scalars are used to address the dark energy problem of late universe \cite{EspositoFarese:2000ij,Clifton:2011jh,Bahamonde:2017ize} as well as the inflationary dynamics of the early universe where nonminimally coupled Higgs \cite{Bezrukov:2007ep} is among the best models to fit observational data \cite{Akrami:2018odb}.

Embracing extra freedom in setting the connection allows to present general relativity in alternative geometric formulations \cite{BeltranJimenez:2019tjy}. By employing teleparallel instead of Riemannian connection the overall curvature vanishes, and by rewriting the Riemann curvature scalar $\lc{R}$ of the Einstein--Hilbert action in terms of the torsion scalar $T$ and a boundary term $B$, $\lc{R}=-T+B$, we obtain the teleparallel equivalent of general relativity \cite{Aldrovandi:2013wha,Krssak:2018ywd} where the boundary term does not contribute to the field equations, and the dynamics of gravity is facilitated by torsion alone. 
Analogously to nonminimal coupling to curvature, it is thus tempting to consider a teleparallel theory where a scalar field is nonminimally coupled to the torsion scalar \cite{Geng:2011aj}. So far, no computation has been performed of quantum corrections to the scalar field on a teleparallel background, while imposing the usual Kaluza--Klein metric ansatz on a five-dimensional teleparallel action yields an effective four-dimensional theory where the scalar field is nonminimally coupled to both the torsion scalar and the boundary term  \cite{Geng:2014yya,Geng:2014nfa}, thus effectively to the curvature scalar $\lc{R}$. But purely heuristically it may still make sense to try to couple the scalar field to the part of the action which encodes the gravitational dynamics, and leave aside the boundary part. 

In the minimal coupling, a scalar field within teleparallel equivalent of general relativity and a scalar field within general relativity behave exactly the same, but introducing a nonminimal coupling to torsion scalar or Riemannian curvature scalar makes a different theory, like the $f(T)$ theories \cite{Ferraro:2006jd,Bengochea:2008gz,Linder:2010py} differ from their $f(R)$ counterparts. For example, the scalar models nonminimally coupled to $T$ do not enjoy invariance under the basic conformal transformations (rescaling of the metric and reparametrization of the scalar field, but leaving the connection intact) unless one introduces an extra coupling to the $B$-term \cite{Yang:2010ji,Bamba:2013jqa,Wright:2016ayu,Hohmann:2018dqh}. Also, in contrast to the case of a scalar nonminimally coupled to curvature, the parameterized post-Newtonian (PPN) calculation tells that a scalar nonminimally coupled to torsion does not give the effective gravitational constant a Yukawa-type correction and the PPN parameters match those of the minimally coupled scalar field \cite{Chen:2014qsa,Sadjadi:2016kwj,Emtsova:2019qsl}. These features may be attributed to the fact that nonminimal coupling to torsion does not give the scalar field equation a matter source term \cite{Jarv:2017npl}. It is possible to write the $f(T)$ action in the scalar-torsion form \cite{Izumi:2013dca}, but the resulting scalar sees no obvious independent dynamics, thus leaving the issue of the number of degrees of freedom a rather puzzling topic \cite{Li:2011rn,Ferraro:2018tpu,Ferraro:2018axk,Ferraro:2020tqk,Blagojevic:2020dyq,Blixt:2020ekl}. 


Usually teleparallel theories are discussed in the local frame field language where the fundamental variables are tetrad and spin connection components, but a more conventional formulation in terms of the metric and affine connection is also possible \cite{BeltranJimenez:2018vdo}. 
In either case, in the nonminimal teleparallel context one must make sure the non-Riemannian part of the connection satisfies its own field equations, and it is better to work in the covariant formulation \cite{Hohmann:2018rwf}, or otherwise the theory would face the embarrassment of lacking local Lorentz invariance (the same applies to $f(T)$ theories \cite{Krssak:2015oua,Golovnev:2017dox}). 
Secondly, one should note that imposing a symmetry on the metric does not immediately extend the same symmetry on the teleparallel connection (as it would in the Riemannian case), but the connection must meet its own symmetry conditions \cite{Hohmann:2019nat}.
For spatially flat Friedmann--Lema\^itre--Robertson--Walker (FLRW) spacetimes it is not a big issue as a simple ansatz automatically does the job, but already for spatially curved cases the situation turns out to be nontrivial \cite{Ferraro:2011us,Tamanini:2012hg,Hohmann:2019nat,Hohmann:2020zre}.

From the outset, flat FLRW cosmological models with a scalar nonminimally coupled to torsion seem quite promising, as they show basic agreement with observational data \cite{Geng:2011ka}, can exhibit phantom-divide crossing (to $\mathrm{\weff}<-1$) without making the field a phantom \cite{Geng:2011aj,Xu:2012jf,Jamil:2012vb,Kucukakca:2013mya}, and can dynamically converge to general relativity in the matter and potential dominated regimes \cite{Jarv:2015odu}.
The dynamical systems analyses of the theory have been chiefly motivated by addressing the late universe dark energy era and mainly focusing on the exponential potential \cite{Wei:2011yr,Xu:2012jf,Jamil:2012vb,Otalora:2013tba,Bahamonde:2015hza,Skugoreva:2016bck,Sadjadi:2015fca,MohseniSadjadi:2016ukp,DAgostino:2018ngy,Bahamonde:2018miw,Gonzalez-Espinoza:2020jss,Bahamonde:2020vfj}. For power-law potentials the studies indicate that nonminimal torsion coupling leads to a smaller variety of possible dynamical regimes than nonminimal curvature coupling \cite{Skugoreva:2014ena,Skugoreva:2016bck,Skugoreva:2017fpc}. 
The evolution of cosmological perturbations has been studied in Refs.\ \cite{Geng:2012vn,Wu:2016dkt,DAgostino:2018ngy,Abedi:2018lkr,Golovnev:2018wbh,Raatikainen:2019qey,Gonzalez-Espinoza:2019ajd,Gonzalez-Espinoza:2021mwr}, but mostly in a Lorentz noncovariant setting.

The aim of the present paper is to use the methods of dynamical systems \cite{Bahamonde:2017ize} to study the cosmological evolution of scalar field models nonminimally coupled to torsion in spatially flat FLRW backgrounds following the approach of Ref.\ \cite{Jarv:2021qpp}, and compare the results with models nonminimally coupled to curvature. The adopted variables $(\phi,\tfrac{\dot{\phi}}{H})$ have the benefit of clearly distinguishing all asymptotic states, and we confirm the insight about the central role of heteroclinic orbits in the phase space for the realization of inflation \cite{Felder:2002jk,UrenaLopez:2011ur,Alho:2014fha,Alho:2015cza,Alho:2017opd,Jarv:2021qpp} in the torsion case as well. Taking positive quadratic nonminimal couplings with quadratic and quartic potentials as examples, we plot the local and global phase portraits, indicate the leading slow roll trajectory of inflation, and mark the path of the last 50 e-folds of accelerated expansion as well as the range of initial conditions leading to it. We observe how turning on nonminimal coupling to torsion has qualitatively different effects compared to the nonminimal curvature coupling models. The unveiled picture conforms with the asymptotic regimes described in Ref.\ \cite{Skugoreva:2014ena} and corroborates with the features found earlier in the perturbation analyses, namely that quadratic potential models require very small torsion coupling to be viable \cite{Gonzalez-Espinoza:2019ajd}, while quartic potentials are problematic in giving successful inflation at all \cite{Raatikainen:2019qey}.
One should also note that the equations of flat FLRW cosmology with nonminimal coupling to torsion scalar are identical to those with nonminimal coupling to nonmetricity scalar \cite{Jarv:2018bgs}, hence our results pertain also to the  models that stem from the yet another alternative geometric formulation of general relativity \cite{BeltranJimenez:2017tkd,Jarv:2018bgs,BeltranJimenez:2019tjy} as well.

The structure of the paper is as follows. In the next Sec. \ref{sec: st cosmology} we compare the cosmological equations of a scalar field nonminimally coupled to torsion and curvature, and explain the basic behavior in terms of the ``effective mass'' and ``effective potential.'' In Sec.\ \ref{sec: dyn sys} we write the torsion case equations as a dynamical system to prepare for the analyses of Sec.\ \ref{sec: quadratic} on the quadratic potential and Sec.\ \ref{sec: quartic} on the quartic potential case. We offer the concluding comments in Sec.\ \ref{sec: conclusions}.


\section{Scalar-curvature vs. scalar-torsion cosmology}
\label{sec: st cosmology}

\subsection{Action and cosmological equations}

We can write the action for a scalar field nonminimally coupled to curvature as \cite{EspositoFarese:2000ij,Bezrukov:2007ep}
\begin{equation}
\label{scalarcurvature_action}
S = \tfrac{1}{2}\int d^4x\sqrt{-g}\left\lbrace -F(\phi) R + \partial_\mu \phi \, \partial^\mu \phi - 2{V}(\phi)\right\rbrace 
\,
\end{equation}
using the natural units where the reduced Planck mass $M_{\mathrm{Pl}}=(8\pi G_N)^{-1/2} =1$. The nonminimal coupling function $F(\phi)$ makes the effective Planck mass and correspondingly the effective gravitational ``constant'' dynamical, while we assume $F>0$. We also assume the potential is nowhere negative $V(\phi)\geq 0$, and the scalar is not a ghost, 
\begin{equation}
\label{eq: no ghosts}
    E(\phi)=2F(\phi)+3F_{,\phi}^2 \geq 0 \,.
\end{equation}
Here and below the comma denotes a derivative with respect to the scalar field, e.g.\ $F_{,\phi\phi}=\tfrac{d^2 F}{d \phi^2}$. The condition \eqref{eq: no ghosts} can be observed as giving the correct sign of the scalar kinetic term in the conformally transformed Einstein frame where the dynamics of the metric and the scalar field are explicitly decoupled \cite{Clifton:2011jh,Capozziello:2011et}.

Analogously, the action for a scalar field with nonminimal coupling to torsion is taken to be \cite{Geng:2011aj,Otalora:2013tba,Hohmann:2018rwf}
\begin{equation}
\label{scalartorsion_action}
S = \tfrac{1}{2}\int d^4x\sqrt{-g}\left\lbrace F(\phi) T + \partial_\mu \phi \, \partial^\mu \phi - 2{V}(\phi)\right\rbrace 
\,.
\end{equation}
It too exhibits dynamical effective Planck mass and variable gravitational ``constant'', hence again it makes sense to assume $F>0$. The situation about ghosts is not immediately obvious, however, since although a conformal transformation of the metric can resolve the $FT$ term in the action, it will not achieve explicit decoupling, since another term arises which couples the scalar to the torsional boundary term \cite{Yang:2010ji,Bamba:2013jqa,Wright:2016ayu,Hohmann:2018dqh}.

In a flat FLRW background, 
\begin{equation}
\label{eq: FLRW}
    ds^2=dt^2-a^2(t) \, d\mathbf{x}^2 \,,
\end{equation}
the scalar-curvature cosmological equations are in terms of the Hubble function $H=\tfrac{\dot{a}}{a}$ given by \cite{EspositoFarese:2000ij}
\begin{eqnarray}
\label{eq: SC Friedmann1}
3 F H^2 &=& \tfrac{\dot{\phi}^2}{2} +V \,\, {-3 F_{,\phi} H \dot{\phi}} \\
\label{eq: SC Friedmann2}
-2 F \dot{H} &=& \dot{\phi}^2 \,\, {+ F_{,\phi\phi} \dot{\phi}^2 - F_{,\phi} H \dot{\phi} + F_{,\phi} \ddot{\phi}} \\
\label{eq: SC scalar equation}
\ddot{\phi} + 3 H \dot{\phi} &=& - V_{,\phi} \,\, {+ 3 F_{,\phi} (2 H^2 + \dot{H})} \,.
\end{eqnarray}
In the scalar-torsion theory one needs to supplement the metric ansatz \eqref{eq: FLRW} with the corresponding ansatz on  teleparallel connection, which must be endowed with vanishing curvature, obey the same symmetry of spatial homogeneity and isotropy as the metric  \cite{Hohmann:2019nat}, and also satisfy its own antisymmetric field equations \cite{Hohmann:2018rwf}. Fortunately in the present case of Cartesian coordinates \eqref{eq: FLRW} such connection is easy to find \cite{Hohmann:2018rwf,Hohmann:2019nat} and the resulting cosmological equations are \cite{Geng:2011aj,Otalora:2013tba,Hohmann:2018rwf}
\begin{eqnarray}
\label{eq: ST Friedmann1}
3 F H^2 &=& \tfrac{\dot{\phi}^2}{2} + V \\
\label{eq: ST Friedmann2}
-2 F \dot{H} &=& \dot{\phi}^2 \,\, {+ 2 F_{,\phi} H \dot{\phi}} \\
\label{eq: ST scalar equation}
\ddot{\phi} + 3 H \dot{\phi} &=& - V_{,\phi} \,\, {-3 F_{,\phi} H^2} \,.
\end{eqnarray}
Note that in the minimally coupled limit, $F=1$, these two sets of equations \eqref{eq: SC Friedmann1}--\eqref{eq: SC scalar equation} and \eqref{eq: ST Friedmann1}--\eqref{eq: ST scalar equation} are identical as expected, but switching on nonminimal coupling to curvature or torsion introduces differences. Just by looking one gets an impression that among the two, the nonminimal coupling to torsion seems to have less impact, since the Friedmann equation \eqref{eq: ST Friedmann1} stays the same as in the minimal coupling. Also note that in the scalar field equations \eqref{eq: SC scalar equation} and \eqref{eq: ST scalar equation} the $H^2$ term comes with a different sign, which does affect the existence conditions of a regime where $\phi$ is stable \cite{Skugoreva:2014ena}.

The rate of expansion can be conveniently expressed by the effective barotropic index, which for the scalar-curvature case reads
\begin{align}
\weff &= -1 - \frac{2 \dot{H}}{3 H^2} 
= -1 + \frac{2 F_{,\phi\phi} \dot{\phi}^2 - 2 F_{,\phi} H \dot{\phi} + 2 F_{,\phi} \ddot{\phi} + 2 \dot{\phi}^2}{\dot{\phi}^2 -6 F_{,\phi} H \dot{\phi} + 2 V} \,,
\label{eq: SC weff}
\end{align}
while for the scalar-torsion case
\begin{align}
\weff &= -1 - \frac{2 \dot{H}}{3 H^2} 
= -1 + \frac{ 4 F_{,\phi} H \dot{\phi}+2 \dot{\phi}^2}{\dot{\phi}^2+2V} \,.
\label{eq: ST weff}
\end{align}
In the minimally coupled case the barotropic index is bounded by $-1\leq \weff \leq 1$, but in the nonminimally coupled models it is also possible to encounter superaccelerated ($\weff<-1$) as well as superstiff ($\weff>1$) expansion. 
When the scalar field stops, $\dot{\phi}=0, \ddot{\phi}=0$, we have de Sitter like behavior with $\weff=-1$, provided $V>0$.

\subsection{“Effective potential” and “effective mass”}

By manipulating (``debraiding'') the field equations we can remove $\dot{H}$ from the scalar field equation, yielding in the scalar-curvature case
\begin{equation}
\label{eq: SC phi equation}
\ddot{\phi} = \frac{1}{E} (-2 F V_{,\phi} + 4 F_{,\phi} V) -
\frac{\dot{\phi}^2}{E} (3 F_{,\phi\phi} F_{,\phi} + F_{,\phi}) - 3 \dot{\phi} H \,.
\end{equation}
In the last term we could have also replaced $H$ by a cumbersome square root coming from Eq.\ \eqref{eq: SC Friedmann1}, but let it remain for the moment. In the  scalar-torsion case the analogous manipulation gives
\begin{equation}
\label{eq: ST phi equation}
\ddot{\phi} = \frac{1}{F} (- F V_{,\phi} - F_{,\phi} V) - \frac{\dot{\phi}^2}{2F} F_{,\phi} -3 \dot{\phi} H  \,.
\end{equation}
These equations coincide in the minimally coupled limit,
\begin{equation}
\label{eq: minimal coupling phi equation}
\ddot{\phi} = - V_{,\phi} -3 \dot{\phi} H \,.
\end{equation}

To understand the scalar field dynamics it is instructive to compare these equations with a ``mechanical analogue'' of a point particle moving in the $x$-dimension and experiencing a force described by the gradient of an external potential $V$, as well as friction proportional to its velocity,
\begin{equation}
\ddot{x} = \frac{1}{m} (- V_{,x} ) - \dot{x} \, \textrm{(friction)} \,.
\end{equation}
The formal similarity allows us to speak of an ``effective potential'' and ``effective mass'' of the system,\footnote{Another approach with different definitions can be found in Refs.\ \cite{Sadjadi:2015fca,MohseniSadjadi:2016ukp}.} see Table \ref{tab: comparisontable} for the expressions. Roughly speaking, the system finds the points of stationarity, i.e.\ the fixed points, at the extrema of the ``effective potential'' which correspond to the vanishing of the ``effective force''. These points are stable or unstable depending on whether they are located at the local minimum or maximum of the ``effective potential'', respectively. The friction term depends on the effective ``speed'' and does not directly contribute to the existence of fixed points, but it should carry the correct sign to avoid a spontaneous blow-up. 
\begin{table}
    \centering
\begin{tabular}{lcccc}
& Effective mass & Effective potential & Fixed point condition & Stability condition \\
     & $m_{\mathrm{eff}}$ & $V_{\mathrm{eff}}$ & & 
     \\  
&&&&\\
Minimally coupled & $1$ & $V$ & $\frac{V_{,\phi}}{V}=0$ & $\frac{V_{,\phi\phi}}{V} >0$ \\
&&&& \\
Scalar-curvature & $\frac{E}{F^3}$ & $\frac{V}{F^2}$ & $\frac{1}{m_{\mathrm{eff}}} \frac{V_{\mathrm{eff},\phi}}{V}=0$ & $\frac{1}{m_{\mathrm{eff}}} \frac{V_{\mathrm{eff},\phi\phi}}{V}>0$ \\
&&\\
Scalar-torsion & $F$ & $FV$ & $\frac{1}{m_{\mathrm{eff}}} \frac{V_{\mathrm{eff},\phi}}{V}=0$ & $\frac{1}{m_{\mathrm{eff}}} \frac{V_{\mathrm{eff},\phi\phi}}{V}>0$ \\
&&&&\\
\end{tabular}
    \caption{Mechanical analogue quantities of the scalar field cosmological equation in the case of minimal coupling \eqref{eq: minimal coupling phi equation}, nonminimal coupling to curvature \eqref{eq: SC phi equation}, and nonminimal coupling to torsion \eqref{eq: ST phi equation}. The conditions for the fixed points and their stability were deduced in Ref.\ \cite{Dutta:2020uha} for the scalar-curvature case, and follow from Eqs.\ \eqref{eq: dynsys fixedpoint condition}, \eqref{eq: dynsys fixedpoint eigenvalues} in the scalar-torsion case.}
    \label{tab: comparisontable}
\end{table}

The idea of an ``effective potential'' has been advocated as a useful concept in the nonminimal curvature coupling case \cite{Chiba:2008ia,Skugoreva:2014gka,Jarv:2016sow,Dutta:2020uha} where it also turns out to be an invariant quantity under conformal transformations \cite{Jarv:2014hma} and thus rules the dynamics in a frame independent way. (Curiously, it can be also extended to theories with an additional coupling to a Gauss--Bonnet term \cite{Pozdeeva:2019agu,Vernov:2021hxo}.)
In the torsion coupling case the ``effective potential'' was introduced in Ref.\  \cite{Skugoreva:2016bck} to explain the existence and stability of de Sitter fixed points, and noted how these points correspond to the ``balanced solutions'' of Ref.\ \cite{Jarv:2015odu}. In particular, the form of the ``effective potential'' dictates that for the nonminimal couplings $F=1+\xi \phi^N$ and power-law potentials $V \sim \phi^n$, regular de Sitter fixed points exist only for negative $n$ with $0 < -n < N$ \cite{Skugoreva:2016bck}. The different functional form of the ``effective potential'' can also explain the markedly different cosmological behaviors of scalar-curvature and scalar-torsion models \cite{Skugoreva:2016bck}.

Here we would like to complement these insights about the ``effective potential'' by also introducing an  ``effective mass'' of the system, which should not be confused with the mass of the scalar particle in the quantum picture. In the nonminimal coupling case the ``effective mass'' is not constant but depends on the value of the scalar field. Therefore the fixed points where the dynamics of the scalar field stops do not only occur at the extrema of the ``effective potential'', but also when the ``effective mass'' becomes infinite (in the language of the mechanical analogue, an infinitely massive object can not be moved by a force that is finite or ``less'' infinite). For instance with the typical nonminimal coupling function $F=1+\xi \phi^N$ this could happen at the asymptotics $|\phi| \to \infty$ when the ``effective potential'' does not have an extremum there, or for $\xi<0$ at a finite value of $\phi$.
Indeed, it is a bit counterintuitive, but even asymptotically diverging power-law potentials $V\sim \phi^n$ (with $n>1$) can feature an asymptotic fixed point if the respective condition in Table \ref{tab: comparisontable} is satisfied. The explicit examples are the $dS_\infty$ point in the minimally coupled and $dS_s$ in the $\xi<0$ nonminimal scalar-curvature case \cite{Dutta:2020uha}. As we will witness below such situation does also occur in the scalar-torsion models. 

To realize inflation by a heteroclinic orbit starting from either a regular or asymptotic fixed point in the phase space \cite{Felder:2002jk,UrenaLopez:2011ur,Alho:2014fha,Alho:2015cza,Alho:2017opd,Jarv:2021qpp}, that point of origin must be a saddle with the eigenvalue corresponding to the outgoing direction having positive real part. This would correspond to a local or asymptotic maximum of the ``effective potential'', or more precisely not fulfilling the respective stability condition in Table \ref{tab: comparisontable}. In principle, the point could be also nonhyperbolic with the real part of the corresponding eigenvalue zero, but in such case the higher corrections must still make the direction repulsive, although one needs to resort to center manifold theory to carry out a precise analysis.

\subsection{Slow roll}

To ensure sufficient measure of inflationary expansion and generate nearly scale invariant spectrum of perturbations, the system must evolve in a nearly de Sitter like regime, which typically implies a slowly varying regime for the scalar field. 
The corresponding slow roll conditions can be read off from the Friedmann equation \eqref{eq: SC Friedmann1} and  the scalar field equation \eqref{eq: SC phi equation} in the scalar-curvature case \cite{Jarv:2021qpp},
\begin{align}
\label{eq: SC slow roll}
3FH^2 \simeq V \,, \qquad 3H\dot{\phi} \simeq -\frac{2F}{E} \left(V_{,\phi} - \frac{2F_{,\phi}V}{F} \right) \,,
\end{align}
and correspondingly from \eqref{eq: ST Friedmann1}, \eqref{eq: ST phi equation} in the teleparallel scalar-torsion case, 
\begin{align}
\label{eq: ST slow roll}
3FH^2 \simeq V \,, \qquad 3H\dot{\phi} \simeq -V_{,\phi}-\frac{F_{,\phi}V}{F} \,,
\end{align}
which matching the result of Ref.\ \cite{Gonzalez-Espinoza:2019ajd}.
Unsurprisingly, in the minimally coupled limit the conditions coincide in the two cases. One may also note in comparison with Eqs.\ \eqref{eq: SC phi equation} and \eqref{eq: ST phi equation} that the slow roll conditon is satisfied exactly at the fixed points where the scalar dynamics stops.

\section{Dynamical system}
\label{sec: dyn sys}

Let us adopt the following dimensionless variables for the dynamical system \cite{Dutta:2020uha,Jarv:2021qpp}
\begin{align}
\label{eq:dynamical variables}
\phi \,, \qquad z=\frac{\dot{\phi}}{H} \,
\end{align}
(also used in Refs.\ \cite{Sadjadi:2015fca,MohseniSadjadi:2016ukp}).
Since the scalar field $\phi$ remains one of the variables, the dynamical system will close for any function $F(\phi), V(\phi)$.
It is useful to measure the expansion of the universe in dimensionless e-folds $N=\ln a$, introducing ${}'=\tfrac{d}{dN} = \tfrac{1}{H}\tfrac{d}{dt}$ as the derivative of dimensionless time. In particular, this implies $\phi'=z$, and also the number of expansion e-folds along a particular trajectory in the phase space can be calculated easily as 
\begin{equation}
N = \int dN = \int H dt = \int \frac{H}{\dot{\phi}} d\phi = \int \frac{1}{z} d\phi \,.
\end{equation}
Here we assume expanding universe, $\dot{H}>0$, and note that $N>0$ for both increasing and decreasing $\phi$, since in the latter case also $z<0$. 

In these variables, the scalar-curvature system \eqref{eq: SC Friedmann1}-\eqref{eq: SC scalar equation} was studied in detail in Refs.\ \cite{Dutta:2020uha, Jarv:2021qpp}, and we will not repeat the details here. In the scalar-torsion case the system \eqref{eq: ST Friedmann1}-\eqref{eq: ST scalar equation} can be represented as 
\begin{eqnarray}
\label{eq: ST dynsys phi}
\phi' &=& z \\
\label{eq: ST dynsys z}
z' &=&  \frac{1}{2F} \left( (z^2 -6F) (F \frac{V_{,\phi}}{V} + z) + 2F_{,\phi}(z^2-3 F) \right)
%
%
\end{eqnarray}

Like in the scalar-curvature case the physical phase space is not spanned by all values of $\phi,z$, but assuming $V\geq 0$ is bounded by the Friedmann constraint. In the scalar-torsion case from \eqref{eq: ST Friedmann1} we get a bound 
\begin{equation}
\label{eq: ST dynsys unphysical}
6F - z^2 \geq 0 \,.
\end{equation}
A possible way to read this bound is by noting that it is exactly satisfied in the ultimate kinetic regime of the scalar field where the contribution of the potential can be neglected in the Friedmann equation.
Analogously, we can express in these variables the scalar-torsion effective barotropic index \eqref{eq: ST weff} as 
\begin{equation}
\label{eq: ST dynsys weff}
    \weff = -1 + \frac{z (z + 2F_{,\phi})}{3F}
\end{equation}
and the slow roll conditions \eqref{eq: ST slow roll} as
\begin{equation}
\label{eq: ST dynsys slowroll}
  z = -\frac{FV_{,\phi}+F_{,\phi}V}{V}
\end{equation}
The latter traces a curve in the phase space. It is not an exact solution (trajectory) itself, but approximates the leading solution that draws all other solutions in the neighbourhood to run closer and closer to it. 
Along the slow roll curve the effective barotropic index is
\begin{equation}
\label{eq: ST slowroll weff}
\weff^{\mathrm{sr}} =
-1 + \frac{F V^2_{,\phi}}{3 V^2} - \frac{F^2_{,\phi}}{3F} \,.
\end{equation}

The regular fixed points arise at whenever the RHS of Eq.\ \eqref{eq: ST dynsys phi}, \eqref{eq: ST dynsys z} vanish, i.e.\ at the values $\phi_*$ satisfying 
\begin{equation}
\label{eq: dynsys fixedpoint condition}
    \left(-\frac{FV_{,\phi}}{V}-F_{,\phi}\right)\Big|_{\phi_*} = 0 \,.
\end{equation}
It is easy to check that the regular fixed points correspond to de Sitter like evolution ($\weff=-1$). They can be either attractors (related to the late universe of dark energy) or saddles (possibly leading to inflation), depending on the signs of the real parts of the eigenvalues 
\begin{equation}
\label{eq: dynsys fixedpoint eigenvalues}
    \lambda = -\frac{3}{2} \pm \frac{3}{2} \sqrt{1 - \frac{4 (F^2 V_{,\phi\phi} + F F_{,\phi \phi} V - 2 F_{,\phi}^2 V)}{3 FV} }
\end{equation}
The properties of the fixed points become clear when analyzed in terms of the effective potential, see the discussion around Table \ref{tab: comparisontable}.

There might be other fixed points at infinity, which are not easily captured by finite analysis. The behavior of the system at $\phi$ infinity can be revealed with the help of Poincar\'e compactification
\begin{align}
\label{eq: Poincare variables}
\phi_p = \frac{\phi}{\sqrt{1+\phi^2+z^2}}\,, \qquad
z_p = \frac{z}{\sqrt{1+\phi^2+z^2}} \,.
\end{align}
The inverse relation between the compact variables $\phi_p, z_p$ and the original ones \eqref{eq:dynamical variables} is
\begin{align}
\label{eq: inverse Poincare}
\phi=\frac{\phi_p}{\sqrt{1-\phi_p^2-z_p^2}} \,, \qquad z=\frac{z_p}{\sqrt{1-\phi_p^2-z_p^2}} \,.
\end{align}
By construction the compact variables are bounded, 
$\phi_p^2+z_p^2\leq 1$, and map an infinite phase plane onto a disc with unit radius. We can express the dynamical system along with the fixed points, slow roll curves, and so on in terms these new variables, to get a compact global picture of the full phase space including the asymptotic regions.

\section{Quadratic potential}
\label{sec: quadratic}

Let us take as the first example 
\begin{equation}
\label{eq: quadratic model}
F = 1 + \xi \phi^2 \,, \qquad V= \frac{m^2}{2} \phi^2 + \Lambda
\end{equation}
and consider only $\xi>0$ to avoid the complications of nonminimal coupling becoming zero and possibly negative.
Like Ref.\ \cite{Jarv:2021qpp} we treat $\Lambda$ as a tiny regularizing parameter, to avoid the dynamical equation \eqref{eq: ST dynsys z} from becoming singular at the origin. We compute all quantities with $\Lambda>0$ and then apply the limit $\Lambda \rightarrow 0$ to present the results. Physically the  $\Lambda$ term can be interpreted as a cosmological constant which is relevant only for the late universe and is many magnitudes smaller that the energy scale of inflation. Since we are only interested in the early universe dynamics and not in the late oscillations of the scalar field around the global minimum of the potential, we are well justified to do it.

\subsection{Finite analysis}

\begin{figure*}
\vspace{-1cm}
	\centering
	\subfigure[]{
		\includegraphics[width=6.8cm]{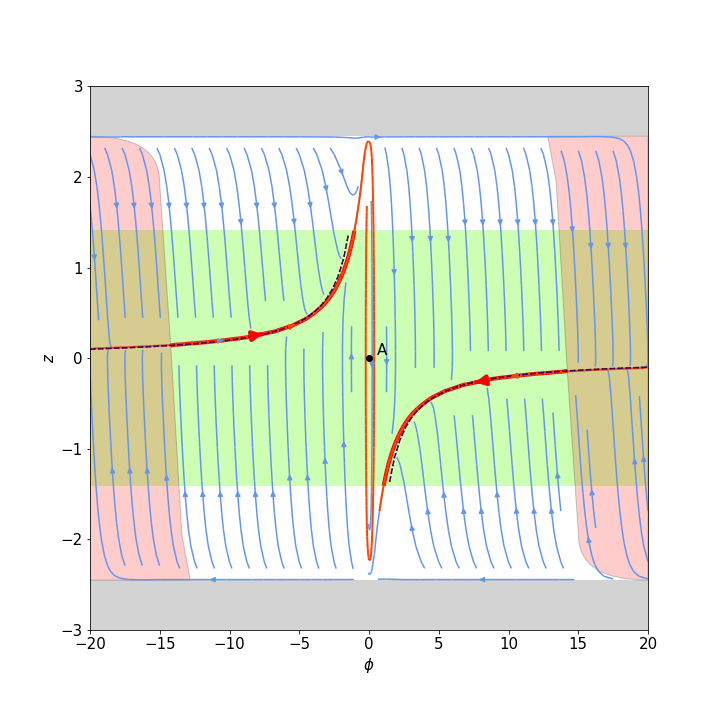} \label{quadratic_xi_0_finite}}
	\qquad
	\subfigure[]{
		\includegraphics[width=6.8cm]{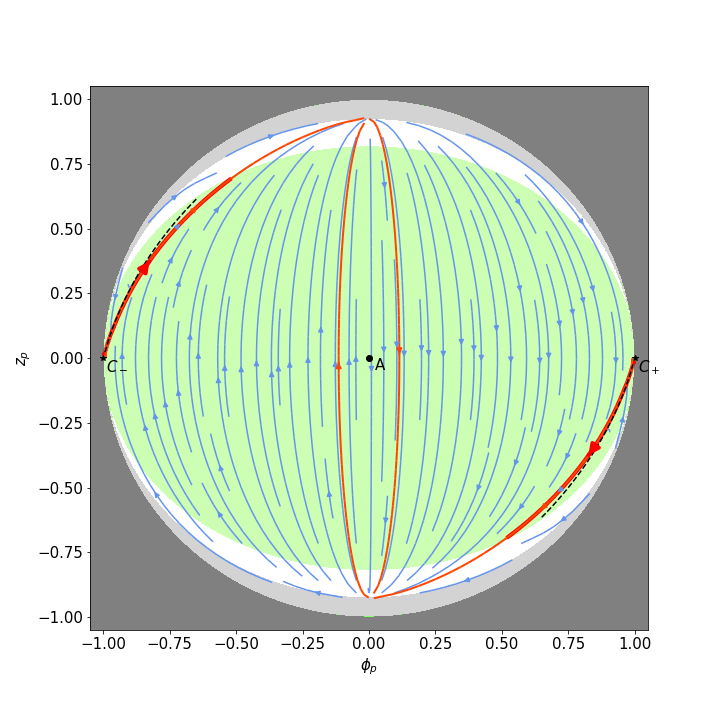} \label{quadratic_xi_0_infinite_full} }
	\\
	\subfigure[]{
		\includegraphics[width=6.8cm]{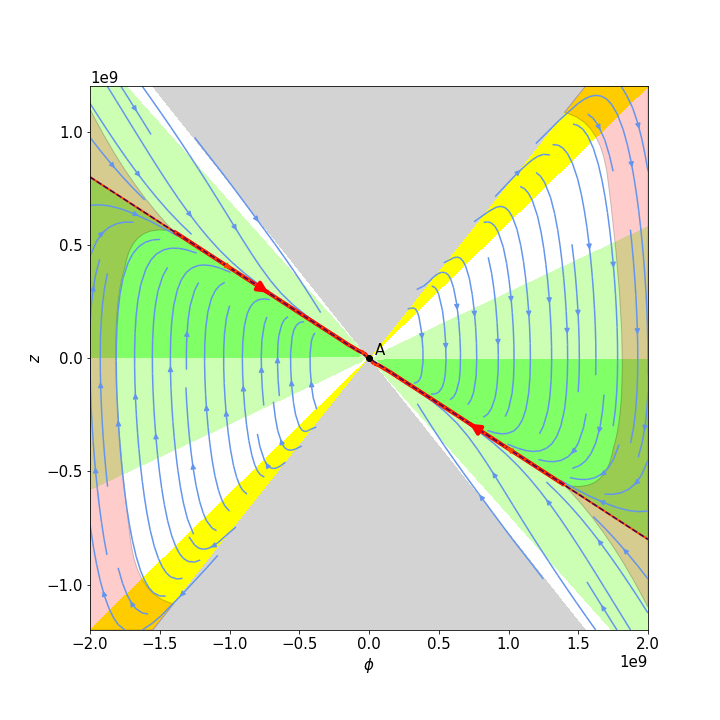} \label{quadratic_xi_0.1_finite}}
	\qquad
	\subfigure[]{
		\includegraphics[width=6.8cm]{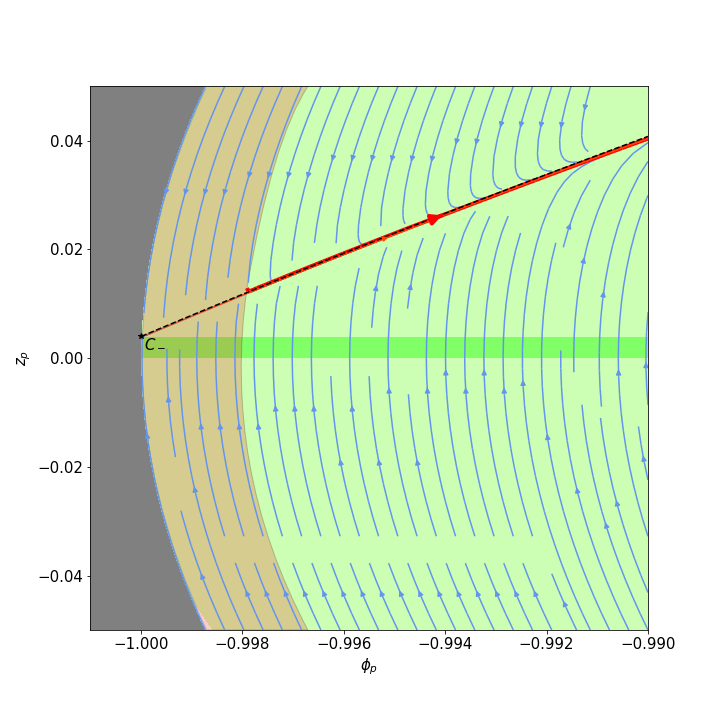} \label{quadratic_xi_0.001_infinite_C_corner} }
	\\
	\subfigure[]{
		\includegraphics[width=6.8cm]{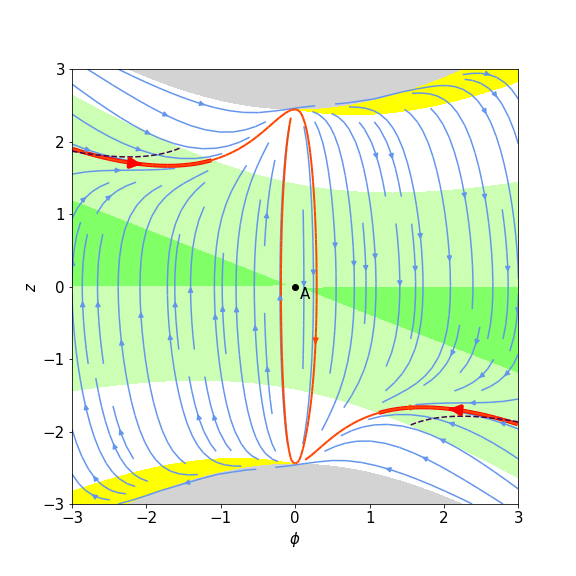} \label{quadratic_xi_0.1_infinite_A}}
	\qquad
	\subfigure[]{
		\includegraphics[width=6.8cm]{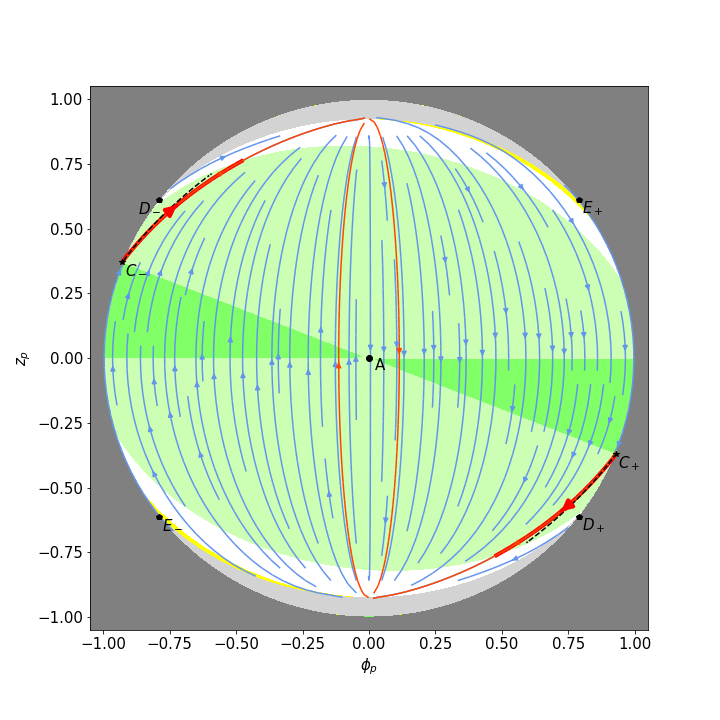} \label{quadratic_xi_0.1_infinite_full}}
	\caption{Cosmological dynamics of the quadratic model \eqref{eq: quadratic model} with $\Lambda=0$, and $\xi=0$ (panels a, b), $\xi=0.001$ (panel d), $\xi=0.1$ (panels c, e, f). Green background stands for superaccelerated, light green accelerated, white decelelerated, and yellow superstiff expansion, while grey covers the unphysical region. Orange trajectories are heteroclinic orbits between the fixed points, wider red part of the master trajectory highlights the 50 last e-folds of inflation, while red semi-transparent cover shows the basin of initial conditions leading to 50 e-folds. The dashed curve marks the path of slow roll approximation.}
\label{fig: portraits quadratic}
\end{figure*}

\begin{figure*}
    \centering
	\subfigure[]{
		\includegraphics[width=6.8cm]{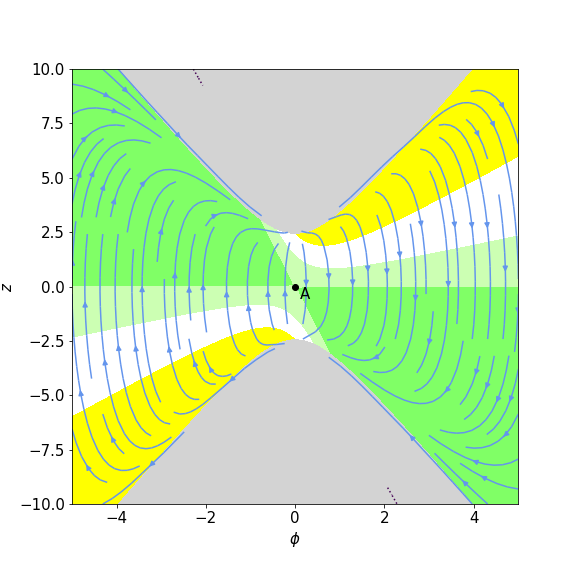} \label{fig: quadratic_xi_1_finite}}
	\qquad
	\subfigure[]{
		\includegraphics[width=6.8cm]{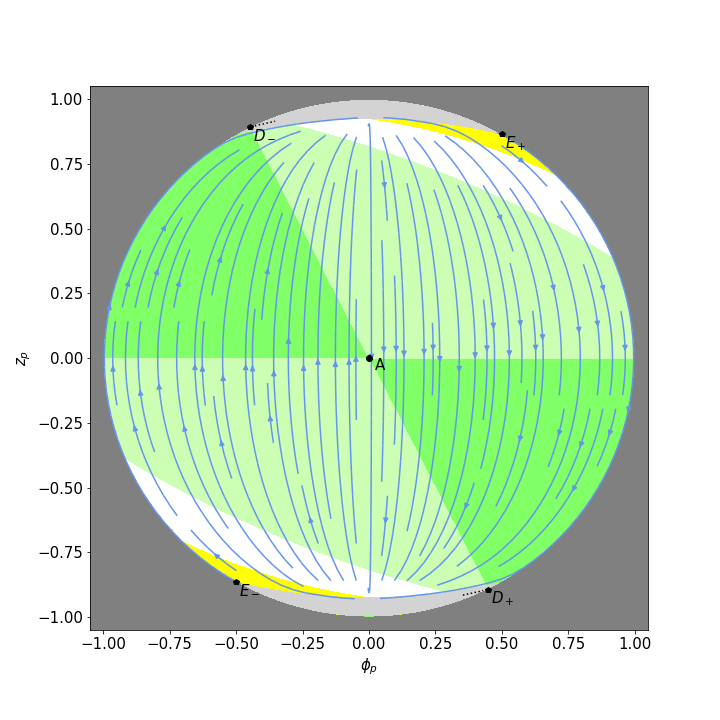} \label{fig: quadratic_xi_0.5_infinite_full}}
	\caption{Cosmological dynamics of the quadratic model \eqref{eq: quadratic model} with $\Lambda=0$, and $\xi=0.5$ (panel b), $\xi=1$ (panel a). Green background stands for superaccelerated, light green accelerated, white decelelerated, and yellow superstiff expansion, while grey covers the unphysical region. Orange trajectories are heteroclinic orbits between the fixed points, wider red part of the master trajectory highlights the 50 last e-folds of inflation, while red semi-transparent cover shows the basin of initial conditions leading to 50 e-folds. The dotted curve marks where the path of the slow roll approximation would be.}
	\label{fig: portraits quadratic 2}
\end{figure*}

For the model \eqref{eq: quadratic model} the dynamical system \eqref{eq: ST dynsys phi}--\eqref{eq: ST dynsys z} is
\begin{align}
\label{eq: dynsys quadratic finite phi}
\phi' &= z \\
\label{eq: dynsys quadratic finite z}
z' &= \frac{\phi(z^2 - 6)(\phi z + 2) + 6\xi \phi^3(z^2 -\phi z - 6)  -24\xi^2 \phi^5}{2\phi^2(1+\xi\phi^2)} \,.
\end{align}
Note that by construction \eqref{eq: ST dynsys z} the factor $m$ cancels out from the dynamical system \eqref{eq: dynsys quadratic finite z}. Also note, that due to the symmetry of the model \eqref{eq: quadratic model} the system \eqref{eq: dynsys quadratic finite phi}--\eqref{eq: dynsys quadratic finite z} is symmetric under the reflection $(\phi \to -\phi, z \to -z)$. The trajectories of the system are depicted on Figs.\ \ref{fig: portraits quadratic} and \ref{fig: portraits quadratic 2}, and the phase space exhibits a ``diagonal'' symmetry.

The barotropic index \eqref{eq: ST dynsys weff} is
\begin{align}
\label{eq: quadratic weff}
\weff &= -1 + \frac{z(z+ 4\xi \phi)}{3(1+ \xi\phi^2)} \,.
\end{align}
On the plots the superaccelerated ($\weff<-1$) part of the phase space is colored green, while acceleration ($-1 \leq \weff <-\tfrac{1}{3}$) is represented by light green, deceleration ($-\tfrac{1}{3} \leq \weff \leq 1$) by white, and the expansion rate corresponding to superstiff equation of state ($\weff>1$) by yellow backgrounds, respectively.

The physical phase space is bounded by \eqref{eq: ST dynsys unphysical}
\begin{align}
\label{eq: quadratic unphysical}
z^2 - 6\xi \phi^2 < 6 \,.
\end{align}
We see that turning on nonminimal coupling extends the physical phase space towards larger $|z|$ for larger $|\phi|$, and infinite $|z|$ become possible at infinite $|\phi|$. One can check explicitly that the solutions do not cross this boundary, e.g.\ by computing the scalar product of the flow vector $(\phi',z')$ and a vector normal to the boundary and seeing it is zero. This also means there is a solution trajectory running along the boundary. On the plots the unphysical regions are covered by grey color.

In the minimally coupled case the layout of the phase space is very simple, as the lines of constant $\weff$ are horizontal, spanning from $\weff=-1$ at $z=0$ to the stiff expansion $\weff=+1$ in the ultimate kinetic regime at the boundary of the physical phase space at $z_b^\pm=\pm\sqrt{6}$.
 Nonminimal coupling distorts this straight picture, the boundary does not correspond to a fixed $\weff$, and superstiff behaviour is also possible near the boundary of the phase space. In particular, tracing the boundary
 \begin{align}
 \label{eq: quadratic z_b}
 z_{b}^\pm (\phi) &= \pm\sqrt{6 + 6\xi\phi^2}
 \end{align}
 where ``$+$'' corresponds to the upper boundary (maximal value of $z$) and ``$-$'' to the lower boundary (minimal value of $z$) to the asymptotics $|\phi|\to \infty$ we get the value of the barotropic index as
 \begin{align}
 \label{eq: quadratic w_eff_b}
\weff_{b}^\pm &= 1 \pm \sqrt{\frac{32 \xi}{3}}
\end{align}
Here $\weff_{b}^-$ corresponds to ``upper left'' $(-\infty,z_b^+)$ and ``lower right'' $(\infty,z_b^-)$ boundary asymptotic, and drops from the value $+1$ of minimal coupling ($\xi=0$) to $-\infty$ for infinite coupling ($\xi\rightarrow \infty$). Analogously, $\weff_{b}^+$ corresponds to ``upper right'' $(\infty,z_b^+)$ and ``lower left'' $(-\infty,z_b^-)$ boundary asymptotic, it increases from the value $+1$ of minimal coupling to infinite value for inifinite coupling. 

The slow roll curve \eqref{eq: ST dynsys slowroll} is given by
\begin{align}
\label{eq: quadratic slow roll curve, z}
z &= -4\phi\xi-\frac{2}{\phi} \,
\end{align}
and is plotted with a dashed black line on the plots. On Fig.\ \ref{fig: portraits quadratic} left column we can observe how it approximates quite well the leading solution (drawn in orange) which attracts the neighbouring solutions. Like in the scalar-curvature case \cite{Jarv:2021qpp}, the approximation curve overestimates $|z|$ compared to the actual attractor solution and thus exits the accelerated expansion regime at slightly higher $|\phi|$. Although the trajectories experience accelerated expansion already while getting closer to the slow roll mode, the expansion during this approach amounts to a few e-folds at best. Only in the vicinity of the leading trajectory the evolution of the scalar field is slow enough to generate significant expansion. On the plots the part of the leading trajectory that corresponds to the last 50 e-folds before the end of accelerated expansion is marked by a red highlight. 

The set of initial conditions which lead to at least 50 e-folds of expansion are covered by a semi-transparent light red hue on the plots. In the discussions of good initial conditions for inflation it is usually assumed that the energy density of the field is Planckian, since one may expect that the effects of quantum gravity will become significant above this scale and alter the dynamics. However, in the nonminimally coupled theories where the effective Planck mass is field dependent, the assessment becomes more involved. For this reason some authors have preferred to investigate the whole range of good initial conditions also looking beyond the Planckian density as measured in the units of late Universe \cite{Mishra:2019ymr}. Here we follow the approach of Ref.\ \cite{Jarv:2021qpp} where the focus turned form good initial conditions to good trajectories, as a single trajectory can represent a set of possible initial data. Looking at the plot  \ref{quadratic_xi_0_finite} of minimal coupling we see that beyond certain $|\phi|$ almost all trajectories converge to the leading trajectory and experience sufficient inflation, although there are also trajectories at high kinetic regime near the physical boundary of the phase space which approach the leading trajectory too late for completing 50 e-folds. Comparing this with Fig.\ \ref{quadratic_xi_0.1_finite} it becomes apparent that increasing $\xi$ pushes the wide zone of good initial conditions back to large $|\phi|$ and thus reduces the likelihood that random initial conditions could give sufficient inflationary expansion (note that on the plot \ref{quadratic_xi_0.1_finite} the units are magnified by $10^9$).

For minimal coupling the slow roll curve starts asymptotically at $(\pm \infty,0)$, but even for a tiny nonminimal coupling the asymptotic starting point jumps to infinite $|z|$. For small $\xi$ this in itself is not a grave problem since the physical phase space also extends to infinite $|z|$ at the asymptotics. Moreover, form Eq.\ \eqref{eq: ST slowroll weff} we see that in the asymptotic limit $\weff^{sr}_{|\phi|\to \infty}=-1$, i.e.\ de Sitter like expansion. However, for $\xi>\tfrac{3}{8}$ the slow roll curve \eqref{eq: quadratic slow roll curve, z} lies entirely in the unphysical region of the phase space, i.e.\ beyond the bound \eqref{eq: quadratic z_b}. Although for larger $\xi$ the phase space is still filled by the zones of accelerated expansion, the scalar field evolves there fast and inflation is not realized. For instance Fig.\ \ref{fig: quadratic_xi_1_finite} shows how in the case of $\xi=1$ the slow roll curve would occur in the unphysical region instead, marked by a dotted line.

For general $\xi$ there are three regular fixed points, however, for $\xi>0$ there is only one point
\begin{align}
    A :& \qquad (0,0) \,.
\end{align}
It is a stable focus and corresponds to the late universe. We can expect more fixed points to reside in the asymptotics, though. Although the potential, effective potential, and their derivatives all diverge as $|\phi| \to \infty$, the fixed point condition listed in Table \ref{tab: comparisontable} is still satisfied in the asymptotic limit. We will investigate the issue in the next subsection.

\subsection{Infinite analysis}

In terms of the compact Poincar\'e variables \eqref{eq: Poincare variables} the system \eqref{eq: dynsys quadratic finite phi}, \eqref{eq: dynsys quadratic finite z} is
\begin{align}
\label{eq: dynsys quadratic infinite phi}
\phi_p' =& \frac{z_p(14q^2 + 6q\phi_p z_p - \phi_p z_p^3)}{2(q + \xi\phi_p^2)} + \frac{\xi \phi_p^2 z_p(19q + 3\phi_p z_p - 2z_p^2)}{q + \xi\phi_p^2} +  \frac{12\xi^2 \phi_p^4 z_p}{q + \xi\phi_p^2} \\
z_p' =&-\frac{12q^3 + 12q^2\phi_p^2 + 6q^2\phi_pz_p - 2q^2z_p^2 + 6q\phi_p^3z_p - q\phi_pz_p^3 - \phi_p^3z_p^3}{2\phi_p(q + \xi\phi_p^2)} \nonumber \\
 & - \frac{ \xi\phi_p(18q^2 + 18q\phi_p^2 + 3q\phi_pz_p - 3qz_p^2 + 3\phi_p^3z_p - 2\phi_p^2z_p^2)}{q + \xi\phi_p^2}
 -\frac{12\xi^2\phi_p^3(q + \phi_p^2)}{q + \xi\phi_p^2} \,.
\label{eq: dynsys quadratic infinite z}
\end{align}
We have introduced a shorthand notation 
\begin{align}
\label{eq: q def}
    q = 1 - \phi_p^2 - z_p^2 \,.
\end{align}
as the asymptotic infinity of the phase space is mapped to $q=0$. The system still retains its ``diagonal'' symmetry, now manifesting in terms of the compact variables as $(\phi_p \to -\phi_p, z_p \to -z_p)$.
In these variables we can also express the effective barotropic index \eqref{eq: quadratic weff} as
\begin{align}
\label{eq: quadratic weff infinite}
\weff &= -1+\frac{z_p\left(z_p+4\xi\phi_p \right)}{3\left(q+\xi\phi_p^2 \right)} \,,
\end{align}
the physical phase space domain \eqref{eq: quadratic unphysical} as
\begin{align}
\label{eq: quadratic unphysical infinite}
\frac{z_p^2-6\xi \phi_p^2}{q} < 6 \,,
\end{align}
and the slow roll curve \eqref{eq: quadratic slow roll curve, z} as 
\begin{align}
\label{eq: quadratic slow roll curve infinity}
\frac{2q +\phi_pz_p+4\xi\phi_p^2}{\sqrt{q}\phi_p} &= 0 \,.
\end{align}
The final attractor fixed point of the late universe is mapped to the origin,
\begin{align}
    A :& \qquad (0,0) \,,
\end{align}
in these variables, and retains its focus character too.

By carefully taking the $q\to 0$ limit in the system \eqref{eq: dynsys quadratic infinite phi}--\eqref{eq: dynsys quadratic infinite z} we can detect an asymptotic pair of fixed points
\begin{align}
    C_\pm :& \qquad (\pm \Omega_1, \mp \sqrt{1-\Omega_1^2}) \,,
\end{align}
where
\begin{align}
    \Omega_1 &= \frac{1}{\sqrt{1+16 \xi^2}} \,.
    \label{eq: Omega_1}
\end{align}
Checking with Eq.\ \eqref{eq: quadratic weff infinite}, these points represent de Sitter like expansions,  $\weff=-1$. One should note that in the nonminimal coupling these points do correspond to a regime where the scalar field evolves, as $z$ is not zero. On the other hand, the evolution of the scalar field must be slow, since the slow roll curve \eqref{eq: quadratic slow roll curve infinity} reaches this point in the asymptotic limit. These points perform as saddles, as can be seen most clearly on the plot \ref{quadratic_xi_0.001_infinite_C_corner}. The heteroclinic orbits from $C_\pm$ to $A$ rule the inflationary dynamics, and other trajectories in the phase space are attracted by these leading trajectories. As explained before, if another trajectory manages to get to the vicinity of the heteroclinic orbit at sufficiently high $|\phi|$, it has a chance to experience more than 50 e-folds of accelerated expansion.

In the asymptotics there is also another set of fixed points 
\begin{align}
    D_\pm :& \qquad (\pm \Omega_2, \mp \sqrt{1-\Omega_2^2}) \,, \\
    E_\pm :& \qquad (\pm \Omega_2, \pm \sqrt{1-\Omega_2^2}) \,, 
\end{align}
where
\begin{align}
\label{eq: Omega_2}
    \Omega_2 &= \frac{1}{\sqrt{1+6 \xi}} \,.
\end{align}
These points correspond to the locations where the boundary of the physical phase space \eqref{eq: quadratic unphysical infinite} runs into the asymptotic circle representing infinity. Checking with \eqref{eq: quadratic weff infinite}, the effective barotropic index at these points is indeed given by \eqref{eq: quadratic w_eff_b}, i.e.\ by $\weff^-_{b}$ at $D_\pm$ and by $\weff^+_{b}$ at $E_\pm$. As these points reside at the boundary of the physical phase space, they correspond to the utmost kinetic regime of the scalar field, either rolling ``down'' or ``up'' its potential.
For small $\xi$ the points $D_\pm$ act as  stable nodes where the trajectories originate, but for large $\xi$ they are saddles. The points $E_\pm$ are always saddles. In the limit of minimal coupling, $\xi=0$, the points $D_\pm$ and $E_\pm$ merge with $C_\pm$, giving it a hybrid saddle-node character, as seen on Fig.\ \ref{quadratic_xi_0_infinite_full}.

 By comparing \eqref{eq: Omega_1} and \eqref{eq: Omega_2} it becomes now clear why the model yields no inflation for $\xi>\tfrac{3}{8}$. As the points $C_\pm$ are the origin of the of the leading inflationary trajectories approximated by the slow troll conditions, increasing $\xi$ shifts these points until they move beyond the points $D_\pm$ on the asymptotic circle, whereby they find themselves in the unphysical territory of the phase space. This process can be followed step by step on the Figs.\ \ref{quadratic_xi_0_infinite_full}, \ref{quadratic_xi_0.1_infinite_full}, \ref{fig: quadratic_xi_0.5_infinite_full}. When the points $C_\pm$ have left the physical phase space, the inflationary guiding paths from $C_\pm$ to $A$ are also gone. For the completeness of description let us mention that while the points $C_\pm$ are physical, there are asymptotic heteroclinic orbits running into them from both points $D_\pm$ and $E_\pm$. When the points $C_\pm$ become unphysical, the points $D_\pm$ change their character from an unstable node to a saddle, and there is an asymptotic heteroclinic orbit from $E_\pm$ to $D_\pm$. On the other hand, for any $\xi$ there are  heteroclinic orbits running along the boundary of the physical phase space from $D_\pm$ to $E_\mp$. 

It is fascinating that we can identify our asymptotic fixed points with the corresponding regimes uncovered by a study which used completely different set of dynamical variables. Indeed, our points $D_\pm$ match the properties and the effective barotropic index of points $Q_3$ of Ref.\ \cite{Skugoreva:2014ena}, exhibiting the regime of power-law evolution,
\begin{align}
\label{eq: D solution}
    a(t) &= a_0 |t-t_0|^{\tfrac{1}{3-2\sqrt{6\xi}}} \,, \qquad
    \phi(t) = \phi_0 |t-t_0|^{-\tfrac{\sqrt{6\xi}}{3-2\sqrt{6\xi}}}
\end{align}
(for $\xi\neq \tfrac{3}{8}$), and turning from an unstable node to a saddle at $\xi>\tfrac{3}{8}$, as the eigenvalues tell. Similarly, our points $C_\pm$ match the properties and the effective barotropic index of the points $Q_4$ of Ref.\ \cite{Skugoreva:2014ena}, exhibiting the regime of exponential evolution,
\begin{align}
\label{eq: C solution}
    a(t) &= a_0 e^{\sqrt{\tfrac{m}{2\xi (3-8\xi)}}(t-t_0)} \,, \qquad
    \phi(t) = \phi_0 e^{\sqrt{\tfrac{8m\xi}{ (3-8\xi)}}(t-t_0)}
\end{align}
(for $\xi\neq \tfrac{3}{8}$), and saddle point character, as the eigenvalues tell. The points $Q_1$ and $Q_2$ found in  Ref.\ \cite{Skugoreva:2014ena} occur for negative nonminimal coupling and inverse power law potentials, and thus do not pertain to our model, while our points $E_\pm$ represent a very particular regime and were not found in Ref.\ \cite{Skugoreva:2014ena}.

The global phase portraits are depicted on the right panels of Figs.\ \ref{fig: portraits quadratic}, \ref{fig: portraits quadratic 2} and in summary tell the following story. For small couplings $\xi<\tfrac{3}{8}$ generic solutions start in the ultimate kinetic dominated regime at the points $D_\pm$, but are attracted by the pair of inflationary attractor orbits that originate at the points $C_\pm$, initially manifest slow roll, and end up at the attractor point $A$ of the late universe. Solutions which converge to the leading inflationary trajectory soon enough have a chance to enjoy at least 50 e-folds of accelerated expansion. For larger couplings $\xi\geq \tfrac{3}{8}$ the inflationary solution is not present at all. Then the asymptotic fixed points are all saddles and it is hard to pinpoint the origin of generic solutions, which all exhibit fast oscillations, and end at the point $A$ of the late universe. Therefore this picture is roughly consistent with the perturbation analysis of Ref.\  \cite{Gonzalez-Espinoza:2019ajd}, which found that quadratic potential models require very small torsion coupling to be viable.

\section{Quartic potential}
\label{sec: quartic}

As the second example, let us take the quartic potential,
\begin{equation}
\label{eq: quartic model}
F = 1 + \xi \phi^2 \,, \qquad V= \frac{\lambda}{4} \phi^4 + \Lambda \,,
\end{equation}
and again consider only $\xi>0$ to avoid complications. As before, we will treat $\Lambda$ as a tiny regularizing parameter, applying the limit $\Lambda\to 0$ to present the results.

\subsection{Finite analysis}

\begin{figure*}
\vspace{-1cm}
	\centering
	\subfigure[]{
		\includegraphics[width=6.8cm]{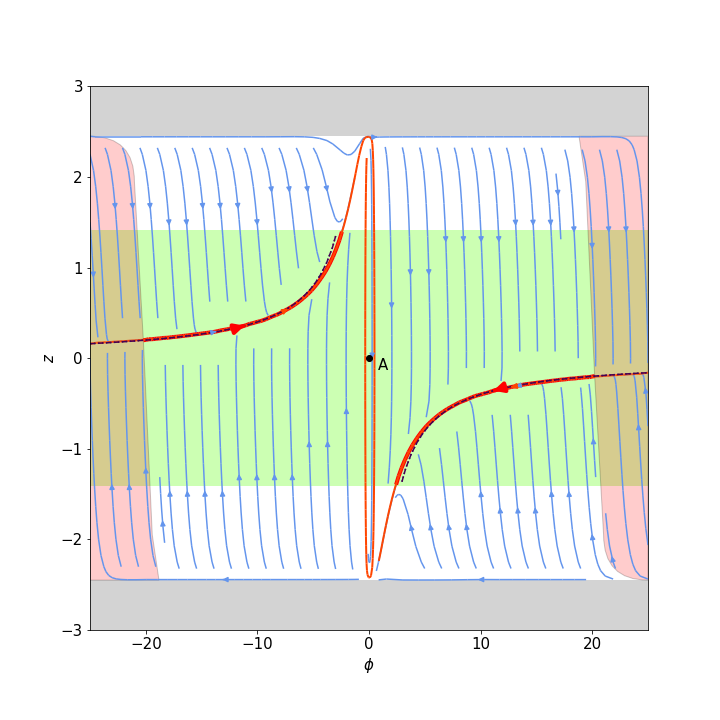} \label{fig: quartic_xi_0_finite}}
	\qquad
	\subfigure[]{
		\includegraphics[width=6.8cm]{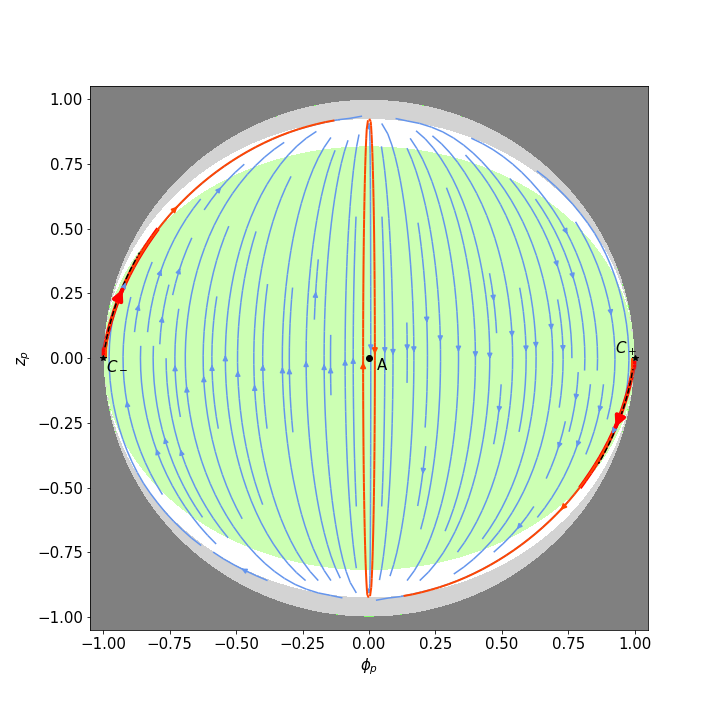} \label{fig: quartic_xi_0_infinite_full} }
	\\
	\subfigure[]{
		\includegraphics[width=6.8cm]{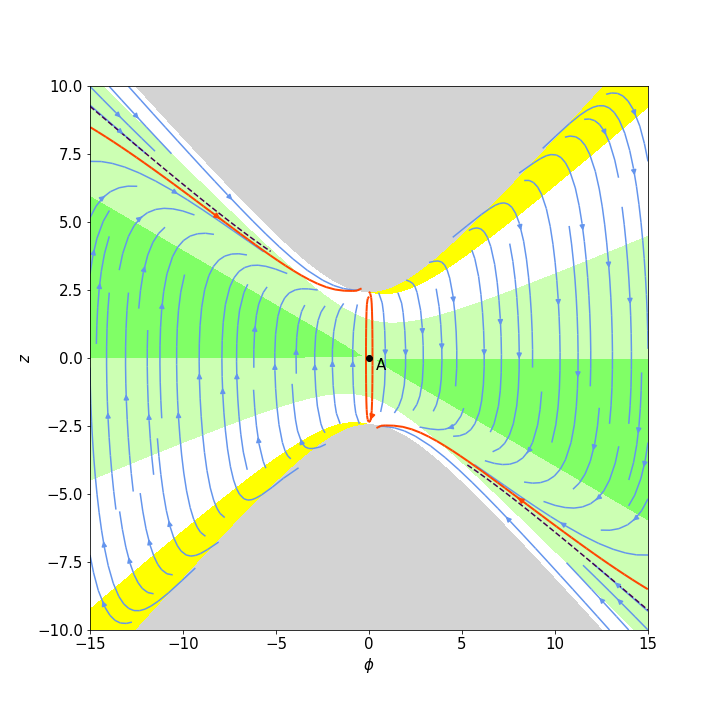} \label{fig: quartic_xi_0.1_finite}}
	\qquad
	\subfigure[]{
		\includegraphics[width=6.8cm]{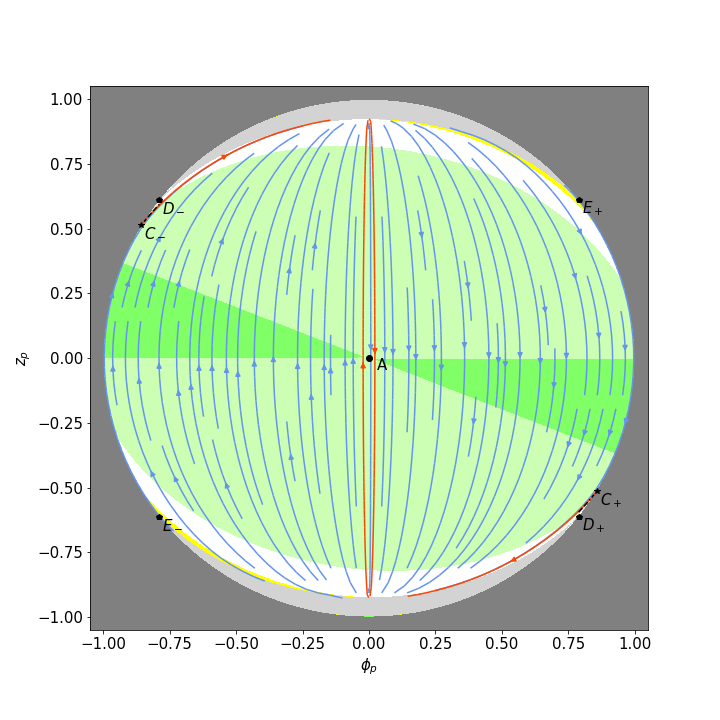} \label{fig: quartic_xi_1_infinite_full} }
	\\
	\subfigure[]{
		\includegraphics[width=6.8cm]{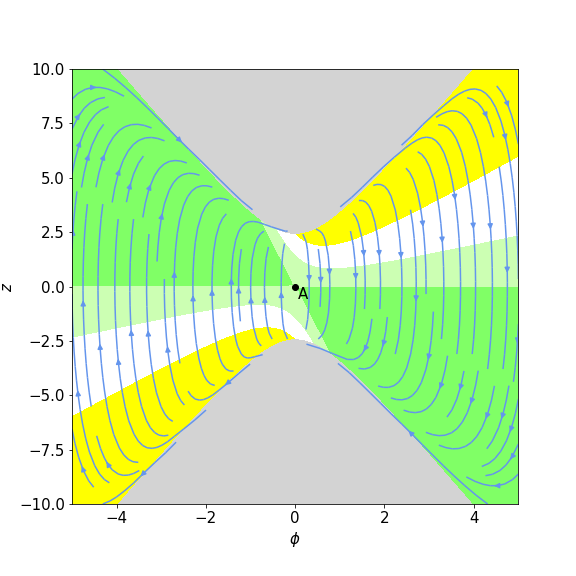} \label{fig: quartic_xi_1_finite}}
	\qquad
	\subfigure[]{
		\includegraphics[width=6.8cm]{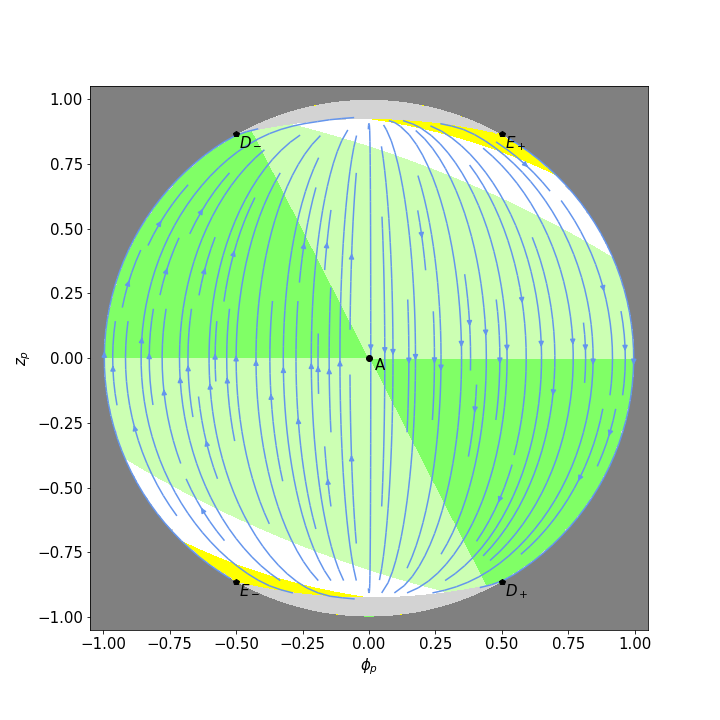} \label{fig: quartic_xi_10_infinite_full}}
	\caption{Cosmological dynamics of the quartic model \eqref{eq: quartic model} with $\Lambda=0$, and $\xi=0$ (panels a, b), $\xi=0.1$ (panels c, d), $\xi=0.5$ (panel f), $\xi = 1$ (panel e). Green background stands for superaccelerated, light green accelerated, white decelelerated, and yellow superstiff expansion, while grey covers the unphysical region. Orange trajectories are heteroclinic orbits between the fixed points, wider red part of the master trajectory highlights the 50 last e-folds of inflation, while red semi-transparent cover shows the basin of initial conditions leading to 50 e-folds. The dashed curve marks the path of slow roll approximation.}
\label{fig: portraits quartic}
\end{figure*}

For the model \eqref{eq: quartic model} the dynamical system \eqref{eq: ST dynsys phi}--\eqref{eq: ST dynsys z} is
\begin{align}
\label{eq: dynsys quartic finite phi}
\phi' &= z \\
\label{eq: dynsys quartic finite z}
z' &= \frac{\phi^3(\phi z+4)(z^2-6) +2\xi \phi^5(4z^2-3\phi z-30)-36\xi^2\phi^7}{2\phi^4 (1+\xi\phi^2)}
\end{align}
Again by construction the factor $\lambda$ cancels out from the dynamical system \eqref{eq: dynsys quartic finite z}. Also, due to the symmetry of the model the system is symmetric under the reflection $(\phi \to -\phi, z \to -z)$, and the phase space exhibits a ``diagonal'' symmetry on Figs.\ \ref{fig: portraits quartic}.

The barotropic index \eqref{eq: ST dynsys weff} is
\begin{align}
\label{eq: quartic weff}
\weff &= -1 + \frac{z(z+ 4\xi \phi)}{3(1+ \xi\phi^2)}
\end{align}
while the respective color coding is the same for superaccelerating, accelerating, usual decelerating and superstiff expansions on Fig.\ \ref{fig: portraits quartic}. 
The physical phase space is bounded by \eqref{eq: ST dynsys unphysical}
\begin{align}
\label{eq: quartic unphysical}
z^2 - 6\xi \phi^2 < 6 \,,
\end{align}
which coincides with the bound of the quadratic model \eqref{eq: quadratic unphysical}, as it is independent of the potential. Therefore the ``upper'' and ``lower'' boundaries of the physical phase space
 \begin{align}
 z_{b}^\pm (\phi) &= \pm\sqrt{6 + 6\xi\phi^2}
 \end{align}
 as well as the effective barotropic index at the respective boundaries,
  \begin{align}
 \label{eq: quartic w_eff_b}
\weff_{b}^\pm &= 1 \pm \sqrt{\frac{32 \xi}{3}}
\end{align}
are exactly the same. Recall that the boundary represents the ultimate kinetic regime of the scalar field, where the effect of the potential can be neglected. 

The slow roll curve \eqref{eq: ST dynsys slowroll} is given by
\begin{align}
\label{eq: quartic slow roll curve, z}
z &= -6\phi\xi -\frac{4}{\phi}
\end{align}
and plotted again with a dashed black line on the plots. On Figs.\ \ref{fig: quartic_xi_0_finite}, \ref{fig: quartic_xi_0.1_finite} we can observe how it approximates the leading solutions (drawn in orange) which attract the neighbouring solutions. Along the leading trajectory and around its immediate vicinity the evolution of the scalar field is the slowest and offers the best conditions for sustained accelerated expansion. However, only for the minimal coupling and really tiny nonminimal coupling we can have at least 50 e-folds of accelerated expansion, marked by a red highlight on the leading trajectory. As before, the set of initial conditions which lead to at least 50 e-folds of expansion are covered by a semi-transparent light red hue on the plots.

For minimal coupling the slow roll curve starts asymptotically at $(\pm \infty,0)$, while after turning on the nonminimal coupling the asymptotic starting point jumps to infinite $|z|$, like we saw already in the quadratic case. But in contrast to the quadratic case, the effective barotropic index along the slow roll curve does not reach the de Sitter value in the asymptotics, showing only
\begin{align}
\label{eq: quartic slowroll weff asymptotic}
    \weff^{sr}_{|\phi| \to \infty} = -1 + 4 \xi
\end{align}
instead. This explains why it is so hard to get sufficient number of e-folds in accelerated expansion for the nonminimal coupling, as even at the beginning of the slow roll path the expansion rate is not high enough. 
Further on, for $\xi>\tfrac{1}{6}$ the slow roll curve \eqref{eq: quartic slow roll curve, z} lies entirely in the unphysical region of the phase space, i.e.\ beyond the bound \eqref{eq: quadratic z_b}, a qualitative change much alike to the quadratic case. Although for larger $\xi$ the phase space is still filled by the zones of accelerated and superaccelerated expansion, the scalar field evolves everywhere fast and inflation is not realized. The plot \ref{fig: quartic_xi_1_finite} illustrates how for larger $\xi$ there is no leading attractor trajectory or slow roll curve to approximate it, while the solutions are instead pushed close to the boundary of the physical phase space, i.e.\ into the strongly kinetic regime.

For $\xi>0$ and $\Lambda\to 0$ the system \eqref{eq: dynsys quartic finite phi}--\eqref{eq: dynsys quartic finite z} has only one regular fixed point,
\begin{align}
    A:& \qquad (0,0) \,
\end{align}
which is a stable node for $\xi\leq \tfrac{3}{8}$ and a stable focus for $\xi > \tfrac{3}{8}$. But the reasoning around Table. \ref{tab: comparisontable} indicates that there should be additional fixed points in the asymptotics, to be revealed in the next subsection.

\subsection{Infinite analysis}

In terms of the compact Poincar\'e variables \eqref{eq: Poincare variables} the system \eqref{eq: dynsys quartic finite phi}, \eqref{eq: dynsys quartic finite z} is
\begin{align}
\label{eq: dynsys quartic infinite phi}
\phi_p' =& \frac{z_p(26q^2 + 6q\phi_p z_p - 2q z_p^2 - \phi_p z_p^3)}{2(q + \xi\phi_p^2)}  + \frac{\xi \phi_p^2 z_p(31q + 3\phi_p z_p - 3z_p^2)}{q + \xi\phi_p^2} +  \frac{18\xi^2 \phi_p^4z_p}{q + \xi\phi_p^2} \\
\label{eq: dynsys quartic infinite z}
z_p' =& -\frac{24q^3 + 24q^2\phi_p^2 + 6q^2\phi_p z_p - 4q^2z_p^2 + 6q\phi_p^3 z_p - 2q\phi_p^2 z_p^2 - q\phi_p z_p^3 - \phi_p^3 z_p^3}{2\phi_p(q + \xi\phi_p^2)} \nonumber \\
& - \frac{\xi\phi_p(30q^2 + 30q\phi_p^2 + 3q\phi_p z_p - 4q z_p^2 + 3\phi_p^3 z_p - 3\phi_p^2 z_p^2)}{q + \xi\phi_p^2} -
\frac{18\xi^2 \phi_p^3(q + \phi_p^2)}{q + \xi\phi_p^2} \,,
\end{align}
where $q$ measures the distance from the circle representing the asymptotics, \eqref{eq: q def}, as before. The system still manifests the ``diagonal'' symmetry $(\phi_p \to -\phi_p, z_p \to -z_p)$ on the left panels of Fig. \ref{fig: portraits quartic}.

In the compact variables we can express the effective barotropic index \eqref{eq: quartic weff} as
\begin{align}
\weff &= -1+\frac{z_p\left(z_p+4\xi\phi_p \right)}{3\left(q+\xi\phi_p^2 \right)} \,,
\end{align}
the physical phase space domain \eqref{eq: quartic unphysical} as
\begin{align}
\label{eq: quartic unphysical infinite}
\frac{z_p^2-6\xi \phi_p^2}{q} < 6 \,,
\end{align}
and the slow roll curve \eqref{eq: quartic slow roll curve, z} as
\begin{align}
\label{eq: quartic slowroll infinite}
\frac{4q +\phi_pz_p + 6 \xi\phi_p^2}{\sqrt{q}\phi_p} &= 0 \,.
\end{align}
The fixed point representing late universe is mapped to the origin,
\begin{align}
    A :& \qquad (0,0) \,,
\end{align}
and retains its characteristics.

By considering the asymptotic $q\to0$ limit of the system \eqref{eq: dynsys quartic infinite phi}--\eqref{eq: dynsys quartic infinite z} we can find an asymptotic pair of fixed points
\begin{align}
    C_\pm :& \qquad \left(\pm \Omega_1, \mp \sqrt{1-\Omega_1^2}\right) \,,
\end{align}
where
\begin{align}
    \Omega_1 &= \frac{1}{\sqrt{1+36 \xi^2}} \,.
\end{align}
Evidently, these points reside at the location where the slow roll curve \eqref{eq: quartic slowroll infinite} reaches the asymptotics. They exhibit expansion rates given by the effective barotropic index \eqref{eq: quartic slowroll weff asymptotic} which is weaker than de Sitter. These points are saddles, and stand as the origin of the heteroclinic orbits which attract other trajectories as they flow to the fixed point $A$ of the late universe. For only extremely small nonminimal couplings $\xi$ there is a limited chance for the surrounding trajectories to converge early enough to the vicinity of these orbits to experience at least 50 e-folds of accelerated expansion.

In the asymptotics one can also find another set fixed points, 
\begin{align}
    D_\pm :& \qquad \left(\pm \Omega_2, \mp \sqrt{1-\Omega_2^2}\right) \,, \\
    E_\pm :& \qquad \left(\pm \Omega_2, \pm \sqrt{1-\Omega_2^2}\right) \,, 
\end{align}
where
\begin{align}
    \Omega_2 &= \frac{1}{\sqrt{1+6 \xi}} \,.
\end{align}
They are situated at the locations where the boundary of the physical phase space \eqref{eq: quartic unphysical infinite} runs into the asymptotic circle. These points function like their counterparts in the quadratic potential case, with the effective barotropic index $\weff^-_{b}$ for $D_\pm$ and $\weff^+_{b}$ for $E_\pm$ given by Eq.\ \eqref{eq: quartic w_eff_b}. 

Also similarly to the quadratic case, the points $C_\pm$ exist only for small nonminimal couplings. At $\xi>\tfrac{1}{6}$ the points $C\pm$ have shifted beyond the points $D_\pm$ and become unphysical. Correspondingly, the points $D_\pm$ are unstable nodes for small nonminimal couplings, but become saddles for $\xi>\tfrac{1}{6}$. The structure of the heteroclinic orbits is the same as in the quadratic case. Incidentally, the critical value $\xi=\tfrac{1}{6}$ corresponds to $\weff^{sr}_{|\phi|\to \infty}=-\tfrac{1}{3}$, i.e.\ where the points $C_\pm$ lose the property of accelerated expanision. 

The fixed points described above match nicely the results of Ref.\ \cite{Skugoreva:2014ena}, where the corresponding regimes we found using a complitely different set of dynamical variables. Our points $D_\pm$ can be identified with the points $Q_3$, exhibiting the regime of power-law evolution \eqref{eq: D solution}. Our points $C_\pm$ can be identified with the points $Q_4$, which in the quartic case do not exhibit exponential evolution \eqref{eq: C solution}, but another power-law regime
\begin{align}
    a(t) &= a_0 |t-t_0|^{\tfrac{1}{6\xi}} \,, \qquad
    \phi(t) = \phi_0 |t-t_0|^{-1} \,.
\end{align}
They are indeed saddles as the eigenvalues tell. Our points $E_\pm$ were not detected in the analysis of Ref.\ \cite{Skugoreva:2014ena}. Note that the variables of Ref.\ \cite{Skugoreva:2014ena} mapped $Q_3$ and $Q_4$ to the same point on the asymptotic circle, which complicated the reading of the respective phase portrait there. In our variables the points $C_\pm$ and $D_\pm$ appeared as distinct for general $\xi$, which concurs with the observation made in Ref.\ \cite{Jarv:2021qpp} that the the set $(\phi,\tfrac{\dot{\phi}}{H})$ is really useful to disentangle different asymptotic power law regimes, and delivers an easy to interpret diagram.

A sample of global phase portraits of the model are shown on the right panels of Fig.\ \ref{fig: portraits quartic}. The story is mostly similar to the quadratic case, but with one important difference. For small couplings $\xi<\tfrac{1}{6}$ the generic solutions start in the ultimate kinetic dominated regime at the points $D_\pm$, but are attracted by the pair of heteroclinic orbits that originate at the points $C_\pm$, may experience slow roll, and end up at the attractor point $A$ of the late universe. Despite the slow roll property, however, the leading solution starts not as an exponential de Sitter, but as a weaker power law. Hence the model is not conductive for proficient inflation even for small nonmininmal coupling. For larger couplings $\xi\geq \tfrac{1}{6}$ the heteroclinic orbit to $A$ is not present at all. Then the asymptotic fixed points are all saddles and it is hard to pinpoint the origin of generic solutions, which all exhibit fast oscillations, and eventually end at the point $A$ of the late universe. Therefore this picture is in rough agreement with the perturbation analysis of Ref.\ \cite{Raatikainen:2019qey}, which found that the nonminimal scalar-torsion Higgs model (dominated by the quartic term in the potential) is unable to produce inflation.

\section{Conclusions}
\label{sec: conclusions}

In this paper we have followed the dynamical systems analysis approach of Ref.\ \cite{Jarv:2021qpp} to study the global behavior of spatially flat FLRW cosmological models with a scalar field nonminimally coupled to torsion in teleparallel setting. Like in the case of nonminimal coupling to curvature, the set of dynamical variables $(\phi,\tfrac{\dot{\phi}}{H}$) offers an easy to interpret picture of global dynamics, and allows to distinguish the different asymptotic regimes as distinct fixed points. We also confirm the key role of heteroclinic orbits, here trajectories from asymptotic saddle fixed points to the final attractor, in the description of inflation. First, these orbits behave as the leading trajectories to which other solutions are attracted to, thus removing the need to fine-tune the initial conditions, as many solutions initially distant in the phase space will soon follow the same path. Second, these solutions trace out the track of slow roll whereby the expansion is nearly de Sitter, thus providing a good venue for generating inflationary phenomenology. Finally, as inflation is described by a trajectory, and not by a fixed point, there is also a graceful exit from the accelerated expansion at a later stretch of that orbit, when it gets close to the final fixed point of late universe. While it is hard to write down the exact analytic form of the inflationary orbit, there is a simple way to extract the correct nonminimal slow roll conditions from the scalar field equations that yield a curve in the phase space which approximates these orbits well until the end of inflation. 
In a broader perspective, the description of inflation is just one aspect of a more comprehensive endeavor to map out the entire history of universe in terms of fixed points and the heteroclinic orbits connecting them \cite{Dutta:2020uha}.

The asymptotic fixed points, from where the inflationary leading trajectory originates, can exhibit de Sitter like expansion ($\weff=-1$), but do not necessarily correspond to a usual de Sitter solution of constant scalar field. In Sec. \ref{sec: st cosmology} we outlined a general heuristic method to detect such fixed points by introducing the concepts of ``effective potential'' and ``effective mass'', inspired by an analogy from mechanics. We point out that the fixed points of the scalar field do not only occur at the extrema of the ``effecive potential'', but can be also caused by the ``effective mass'' becoming infinite. As for scalars nonminimally coupled to curvature and nonminimally coupled to torsion the ``effective potential'' and ``effective mass'' have a rather different functional form, the dynamics of these models can be expected to be rather different, even when the actions look almost identical.

As particular examples we considered models of positive quadratic nonminimal coupling with quadratic and quartic potentials. In the scalar-curvature case the quadratic potential model is endowed with a finite regular saddle de Sitter fixed point which for higher nonminimal coupling shifts closer to the late time attractor and the zone of good initial conditions leading to at least 50 e-folds shrinks to a very narrow stretch \cite{Jarv:2021qpp}. In the scalar-torsion case we found an asymptotic de Sitter point which for higher nonminimal coupling disappears from the physical part of the phase space, making inflation impossible then. Thus although in both cases the most favorable conditions for successful inflation are provided by small nonminimal couplings, the phase portraits are qualitatively different. For quartic potentials, the scalar-curvature case possesses an asymptotic de Sitter point which remains put when the nonminimal coupling increases, but the zone of good initial conditions enlarges, covering almost all available phase space for very large nonminimal coupling \cite{Jarv:2021qpp}. On the other hand, in the scalar-torsion case we found that although the asymptotic saddle fixed point satisfies the slow roll conditions and is the origin of the heteroclinic trajectory, the expansion it corresponds to is power law and not de Sitter like, which severely hampers the prospects for inflation. For larger nonminimal coupling this point also shifts into the unphysical territory of the phase space. Thus the scalar-torsion models with quartic potentials are not really suitable for inflation at all.

One may ask how to improve the chances of inflation for the models with scalar fields nonminimally coupled to torsion? An immediate answer would be to try different potentials and coupling functions. With the heuristic tools of the ``effective potential'' and ``effective mass'' at hand, it should be possible to engineer models behaving like the best models nonminimal coupling to curvature has offered. As the impetus and implications from fundamental physics to the reasonable form of nonminimal coupling to torsion are still unclear, alternatives to the usual quadratic coupling may be well explored. Of course, there are also proposals for more general scalar-torsion theories including an additional coupling to vector torsion or the boundary term \cite{Bamba:2013jqa,Otalora:2014aoa,Bahamonde:2015hza,Gecim:2017hmn}, modification of the kinetic term \cite{Akbarieh:2018oie}, analogue of Horndeski gravity \cite{Bahamonde:2019shr}, and even more general frameworks \cite{Abedi:2018lkr,Hohmann:2018vle,Bahamonde:2020vfj}. The current method of investigation should be suitable for those as well.

\vspace{6pt}

\authorcontributions{Conceptualization, methodology, writing, editing, supervision, funding acquisition, project administration: L.J.; investigation, formal analysis, validation: L.J.\ and J.L.; visualization: J.L.}

\funding{This research was funded by the Estonian Research Council through the project PRG356, as well as by the European Regional Development Fund through the Center of Excellence TK133 “The Dark Side of the Universe”.}

\acknowledgments{L.J.\ is grateful to Alexey Toporensky for useful discussions.}

\conflictsofinterest{The authors declare no conflict of interest.}


\abbreviations{The following abbreviations are used in this manuscript:\\

\noindent
\begin{tabular}{@{}ll}
FLRW & Friedmann--Lema\^itre--Robertson--Walker
\end{tabular}}

\reftitle{References}

\end{document}